\documentclass[aps,prd,reprint,superscriptaddress]{revtex4-1}

\usepackage{graphicx}

\begin{document}


\title{Experimental evidence for the sensitivity of the air-shower radio signal to the longitudinal shower development}

\author{W.D.~Apel}
\affiliation{Karlsruhe Institute of Technology (KIT), Institut f\"ur Kernphysik, Germany}
\author{J.C.~Arteaga}
\affiliation{Universidad Michoacana, Morelia, Mexico}
\author{L.~B\"ahren}
\affiliation{ASTRON, Dwingeloo, The Netherlands}
\author{K.~Bekk}
\affiliation{Karlsruhe Institute of Technology (KIT), Institut f\"ur Kernphysik, Germany}
\author{M.~Bertaina}
\affiliation{Dipartimento di Fisica Generale dell' Universit\`a Torino, Italy}
\author{P.L.~Biermann}
\affiliation{Max-Planck-Institut f\"ur Radioastronomie Bonn, Germany}
\author{J.~Bl\"umer}
\affiliation{Karlsruhe Institute of Technology (KIT), Institut f\"ur Kernphysik, Germany}
\affiliation{Karlsruhe Institute of Technology (KIT), Institut f\"ur Experimentelle Kernphysik, Germany}
\author{H.~Bozdog}
\affiliation{Karlsruhe Institute of Technology (KIT), Institut f\"ur Kernphysik, Germany}
\author{I.M.~Brancus}
\affiliation{National Institute of Physics and Nuclear Engineering, Bucharest, Romania}
\author{P.~Buchholz}
\affiliation{Universit\"at Siegen, Fachbereich Physik, Germany}
\author{E.~Cantoni}
\affiliation{Dipartimento di Fisica Generale dell' Universit\`a Torino, Italy}
\affiliation{INAF Torino, Instituto di Fisica dello Spazio Interplanetario, Italy}
\author{A.~Chiavassa}
\affiliation{Dipartimento di Fisica Generale dell' Universit\`a Torino, Italy}
\author{K.~Daumiller}
\affiliation{Karlsruhe Institute of Technology (KIT), Institut f\"ur Kernphysik, Germany}
\author{V.~de~Souza}
\affiliation{Universidad S\~ao Paulo, Inst. de F\'{\i}sica de S\~ao Carlos, Brasil}
\author{F.~Di~Pierro}
\affiliation{Dipartimento di Fisica Generale dell' Universit\`a Torino, Italy}
\author{P.~Doll}
\affiliation{Karlsruhe Institute of Technology (KIT), Institut f\"ur Kernphysik, Germany}
\author{R.~Engel}
\affiliation{Karlsruhe Institute of Technology (KIT), Institut f\"ur Kernphysik, Germany}
\author{H.~Falcke}
\affiliation{Radboud University Nijmegen, Department of Astrophysics, The Netherlands}
\affiliation{ASTRON, Dwingeloo, The Netherlands}
\affiliation{Max-Planck-Institut f\"ur Radioastronomie Bonn, Germany}
\author{M.~Finger}
\affiliation{Karlsruhe Institute of Technology (KIT), Institut f\"ur Experimentelle Kernphysik, Germany}
\author{B.~Fuchs}
\affiliation{Karlsruhe Institute of Technology (KIT), Institut f\"ur Experimentelle Kernphysik, Germany}
\author{D.~Fuhrmann}
\affiliation{Universit\"at Wuppertal, Fachbereich Physik, Germany}
\author{H.~Gemmeke}
\affiliation{Karlsruhe Institute of Technology (KIT), Institut f\"ur Prozessdatenverarbeitung und Elektronik, Germany}
\author{C.~Grupen}
\affiliation{Universit\"at Siegen, Fachbereich Physik, Germany}
\author{A.~Haungs}
\affiliation{Karlsruhe Institute of Technology (KIT), Institut f\"ur Kernphysik, Germany}
\author{D.~Heck}
\affiliation{Karlsruhe Institute of Technology (KIT), Institut f\"ur Kernphysik, Germany}
\author{J.R.~H\"orandel}
\affiliation{Radboud University Nijmegen, Department of Astrophysics, The Netherlands}
\author{A.~Horneffer}
\affiliation{Max-Planck-Institut f\"ur Radioastronomie Bonn, Germany}
\author{D.~Huber}
\affiliation{Karlsruhe Institute of Technology (KIT), Institut f\"ur Experimentelle Kernphysik, Germany}
\author{T.~Huege}
\affiliation{Karlsruhe Institute of Technology (KIT), Institut f\"ur Kernphysik, Germany}
\author{P.G.~Isar}
\affiliation{Institute for Space Sciences, Bucharest, Romania}
\author{K.-H.~Kampert}
\affiliation{Universit\"at Wuppertal, Fachbereich Physik, Germany}
\author{D.~Kang}
\affiliation{Karlsruhe Institute of Technology (KIT), Institut f\"ur Experimentelle Kernphysik, Germany}
\author{O.~Kr\"omer}
\affiliation{Karlsruhe Institute of Technology (KIT), Institut f\"ur Prozessdatenverarbeitung und Elektronik, Germany}
\author{J.~Kuijpers}
\affiliation{Radboud University Nijmegen, Department of Astrophysics, The Netherlands}
\author{K.~Link}
\affiliation{Karlsruhe Institute of Technology (KIT), Institut f\"ur Experimentelle Kernphysik, Germany}
\author{P.~{\L}uczak}
\affiliation{National Centre for Nuclear Research, Department of Cosmic Ray Physics, {\L}\'{o}d\'{z}, Poland}
\author{M.~Ludwig}
\affiliation{Karlsruhe Institute of Technology (KIT), Institut f\"ur Experimentelle Kernphysik, Germany}
\author{H.J.~Mathes}
\affiliation{Karlsruhe Institute of Technology (KIT), Institut f\"ur Kernphysik, Germany}
\author{M.~Melissas}
\affiliation{Karlsruhe Institute of Technology (KIT), Institut f\"ur Experimentelle Kernphysik, Germany}
\author{C.~Morello}
\affiliation{INAF Torino, Instituto di Fisica dello Spazio Interplanetario, Italy}
\author{J.~Oehlschl\"ager}
\affiliation{Karlsruhe Institute of Technology (KIT), Institut f\"ur Kernphysik, Germany}
\author{N.~Palmieri}
\affiliation{Karlsruhe Institute of Technology (KIT), Institut f\"ur Experimentelle Kernphysik, Germany}
\author{T.~Pierog}
\affiliation{Karlsruhe Institute of Technology (KIT), Institut f\"ur Kernphysik, Germany}
\author{J.~Rautenberg}
\affiliation{Universit\"at Wuppertal, Fachbereich Physik, Germany}
\author{H.~Rebel}
\affiliation{Karlsruhe Institute of Technology (KIT), Institut f\"ur Kernphysik, Germany}
\author{M.~Roth}
\affiliation{Karlsruhe Institute of Technology (KIT), Institut f\"ur Kernphysik, Germany}
\author{C.~R\"uhle}
\affiliation{Karlsruhe Institute of Technology (KIT), Institut f\"ur Prozessdatenverarbeitung und Elektronik, Germany}
\author{A.~Saftoiu}
\affiliation{National Institute of Physics and Nuclear Engineering, Bucharest, Romania}
\author{H.~Schieler}
\affiliation{Karlsruhe Institute of Technology (KIT), Institut f\"ur Kernphysik, Germany}
\author{A.~Schmidt}
\affiliation{Karlsruhe Institute of Technology (KIT), Institut f\"ur Prozessdatenverarbeitung und Elektronik, Germany}
\author{F.G.~Schr\"oder}
\email{frank.schroeder@kit.edu}
\affiliation{Karlsruhe Institute of Technology (KIT), Institut f\"ur Kernphysik, Germany}
\author{O.~Sima}
\affiliation{University of Bucharest, Department of Physics, Romania}
\author{G.~Toma}
\affiliation{National Institute of Physics and Nuclear Engineering, Bucharest, Romania}
\author{G.C.~Trinchero}
\affiliation{INAF Torino, Instituto di Fisica dello Spazio Interplanetario, Italy}
\author{A.~Weindl}
\affiliation{Karlsruhe Institute of Technology (KIT), Institut f\"ur Kernphysik, Germany}
\author{J.~Wochele}
\affiliation{Karlsruhe Institute of Technology (KIT), Institut f\"ur Kernphysik, Germany}
\author{M.~Wommer}
\affiliation{Karlsruhe Institute of Technology (KIT), Institut f\"ur Kernphysik, Germany}
\author{J.~Zabierowski}
\affiliation{National Centre for Nuclear Research, Department of Cosmic Ray Physics, {\L}\'{o}d\'{z}, Poland}
\author{J.A.~Zensus}
\affiliation{Max-Planck-Institut f\"ur Radioastronomie Bonn, Germany}

\collaboration{LOPES Collaboration}
\homepage{http://www.astro.ru.nl/lopes}
\noaffiliation

\date{\today}

Accepted by Physical Review D

\begin{abstract}
We observe a correlation between the slope of radio lateral distributions, and the mean muon 
pseudorapidity of 59 individual cosmic-ray-air-shower events. 
The radio lateral distributions are measured with LOPES, a digital radio interferometer co-located 
with the multi-detector-air-shower array KASCADE-Grande, which includes a muon-tracking detector.
The result proves experimentally that radio measurements are sensitive to the longitudinal 
development of cosmic-ray air-showers. 
This is one of the main prerequisites for using radio arrays for ultra-high-energy particle 
physics and astrophysics. 
\end{abstract}


\maketitle
The investigation of ultra-high-energy cosmic rays works at the frontiers of particle physics as well as astrophysics. 
Only cosmic rays allow probing particle physics beyond the energy range of man-made accelerators. 
And only high statistics and accurate measurements of energy, direction and mass of cosmic rays will solve the mystery of their origin, i.e.~how natural accelerators achieve energies beyond $10^{20}\,$eV (see references \cite{Haungs2003, Bluemer2009} for reviews). 
Since ultra-high-energy cosmic rays are too rare to be measured directly, they can only be measured indirectly by observing extensive air-showers of secondary particles. In particular, the longitudinal development of the air-showers is of interest for the two main physics goals: for particle physics, since the shower development depends on the interactions at ultra-high energies, and for astrophysics, since the shower development also depends on the mass and type of the primary cosmic-ray particle.

The longitudinal air-shower development can be studied either with measurements of secondary air-shower particles, or by detecting electromagnetic radiation emitted by the air-shower or the traversed air. A method of the first category is the measurement of secondary muons with a muon-tracking detector (MTD), like it is performed at the KASCADE-Grande experiment \cite{Apel2010KASCADEGrande, Doll2002}: the directions of the muon tracks relative to the direction of the air-shower axis carry information on the longitudinal shower development \cite{2011ApelMuonProductionHeights, ZabierowskiIcrc2009}. Currently, the most sophisticated methods for the longitudinal air-shower development fall into the second category. They are based on the detection of fluorescence or Cherenkov light emitted by air-showers, which both is limited to dark, clear and moonless nights -- a disadvantage which can be overcome by measuring the radio emission of air-showers. Digital radio-antenna arrays can measure under almost any conditions and would increase the statistics by a factor of roughly 10 \cite{FalckeNature2005}.

Until now, the sensitivity of radio measurements to the longitudinal air-shower development was predicted by simulations of the radio emission \cite{HuegeUlrichEngel2008}, but the experimental proof was still missing. 
We supply this proof by comparing the mean pseudorapidities of muons measured by KASCADE-Grande with the slope of the radio lateral distribution measured with LOPES \cite{FalckeNature2005, 2010ApelLOPESlateral, SchroederThesis2011}, a digital radio-antenna array co-located with KASCADE-Grande. 
LOPES features a precise time \cite{SchroederTimeCalibration2010} and absolute amplitude calibration \cite{NehlsHakenjosArts2007}, and measures the radio signal between $40$ and $80\,$MHz. Since the radio emission is mainly of geomagnetic origin \cite{Allan1971, FalckeNature2005, Ardouin2009}, the radio signal is predominately east-west polarized. Thus, we restrict this analysis to east-west aligned LOPES antennas, which for most events have a higher measurement quality than north-south aligned antennas.

\begin{figure}
\centering
\includegraphics[width=\columnwidth]{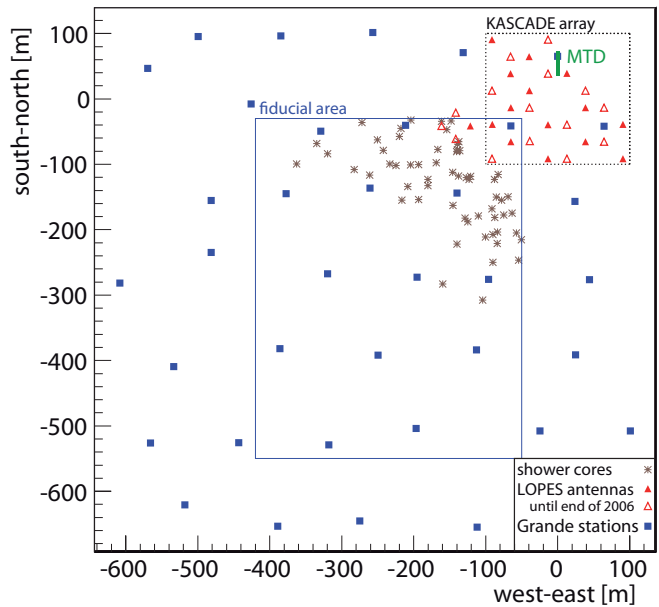}
\caption{Shower cores of the 59 selected events measured simultaneously with the east-west aligned LOPES antennas, the KASCADE-Grande particle-detector array and the muon-tracking detector (MTD). Due to a reconfiguration of LOPES, 15 of the 30 antennas were only available for a part of the events.} \label{fig_eventMap}
\end{figure}

For the analysis we use 59 air-shower events with an estimated primary energy above $5\cdot 10^{16}\,$eV, which were measured between December 2005 and May 2009 with both the MTD of KASCADE-Grande as well as with LOPES (figure \ref{fig_eventMap}). 
The MTD consists of 16 individual muon telescopes with three horizontal layers and one vertical layer of streamer-tube chambers, providing a total detection area of $128\,$m\textsuperscript{2} for vertical muons \cite{Doll2002}. It tracks individual muons, and allows a reconstruction of the radial angle $\rho$, and the tangential angle $\tau$ to the shower axis \cite{Zabierowski2003} measured by KASCADE-Grande with an angular resolution of about $0.3^\circ$ each. We limit the analysis to events with a distance between the MTD and the air-shower axis from $160\,$m to $320\,$m. 
For most of the less distant events the MTD is hit by too many particles to reconstruct the individual muon tracks, and for most of the more distant events the radio signal is too weak. We apply several quality cuts on the KASCADE-Grande reconstruction, e.g., a fiducial area cut to guarantee an accurate reconstruction of the shower geometry and primary energy. Furthermore, we apply several cuts on the LOPES measurement accepting only events with a radio signal clearly distinguishable from noise. Of those events passing these cuts we exclude 1 event recorded during a thunderstorm, and 19 events for which the fit of the exponential-lateral-distribution function fails (see reference \cite{SchroederThesis2011} for a description of the cuts). To eliminate scattered, low-energy muons from the analysis \cite{2011ApelMuonProductionHeights}, we use only muon tracks with $\rho < 8^\circ$ and $\tau < 2^\circ$. Our results are based on the henceforth used 59 events which pass all these cuts. The cuts naturally influence the presented results on a quantitative level (e.g., the numbers resulting from the performed fits), but the qualitative result, namely that there is a correlation between the mean muon pseudorapidity and the slope of the radio lateral distribution, does not depend on the specific values for the cuts.

As observable for the longitudinal development of the air-showers we use the mean pseudorapidity of the muons, where the pseudorapidity of each muon is calculated as $-\ln(\sqrt{\tau^2+\rho^2}/2)$ \cite{Zabierowski2003}. Due to the experimental geometry, i.e.~the confined muon-to-shower-axis distance range, the larger the pseudorapidity of a muon from a given air-shower the larger is the production height along the shower axis.
In principle, the present analysis could be performed directly with the muon-production heights. However, the uncertainty in reconstruction of the distance to the shower axis introduces an additional uncertainty to the production height. Thus, muon pseudorapidities can be measured with a higher precision, and the reported result has a higher significance for the pseudorapidities than for the production heights.
For each shower, we average over the pseudorapidities of all muons (on average $23\pm9$ muon tracks per event). Measured at a certain distance to the shower axis, large mean pseudorapidities mean that the shower development started at a large distance to the detector. For more details, see references \cite{ZabierowskiIcrc2011, DollISVHECRI2010}. The uncertainties of the mean muon pseudorapidities have been estimated by propagating the angular resolution of $0.3^\circ$ of the individual muon tracks. For this we use the bootstrap method, a numerical Monte-Carlo method which takes into account the uncertainties of each individual data point \cite{Efron1993}.

\begin{figure}
\centering
\includegraphics[width=\columnwidth]{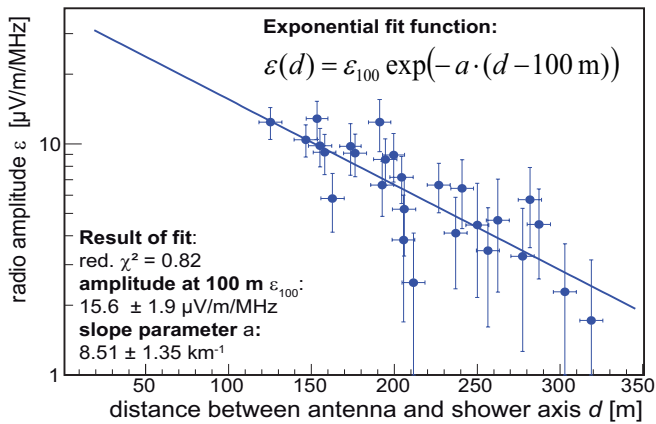}
\caption{Example for a radio lateral distribution measured with LOPES.} \label{fig_radioLDFexample}
\end{figure}

\begin{figure}
\centering
\includegraphics[width=\columnwidth]{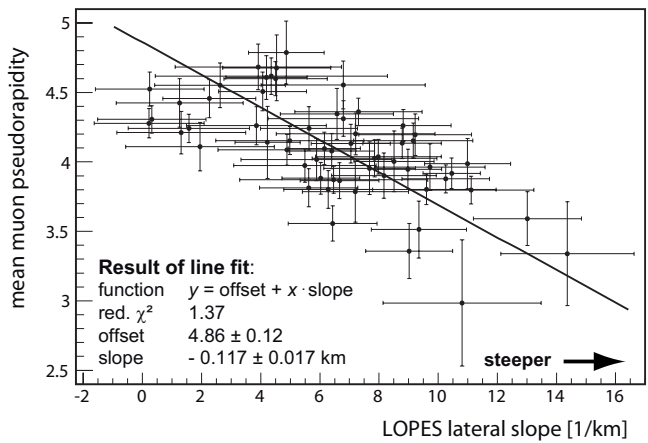}
\caption{Mean muon pseudorapidity and the slope parameter of the radio lateral distribution for the 59 events.} \label{fig_correlationPlot}
\end{figure}

\begin{figure*}[t]
\centering
\includegraphics[width=\columnwidth]{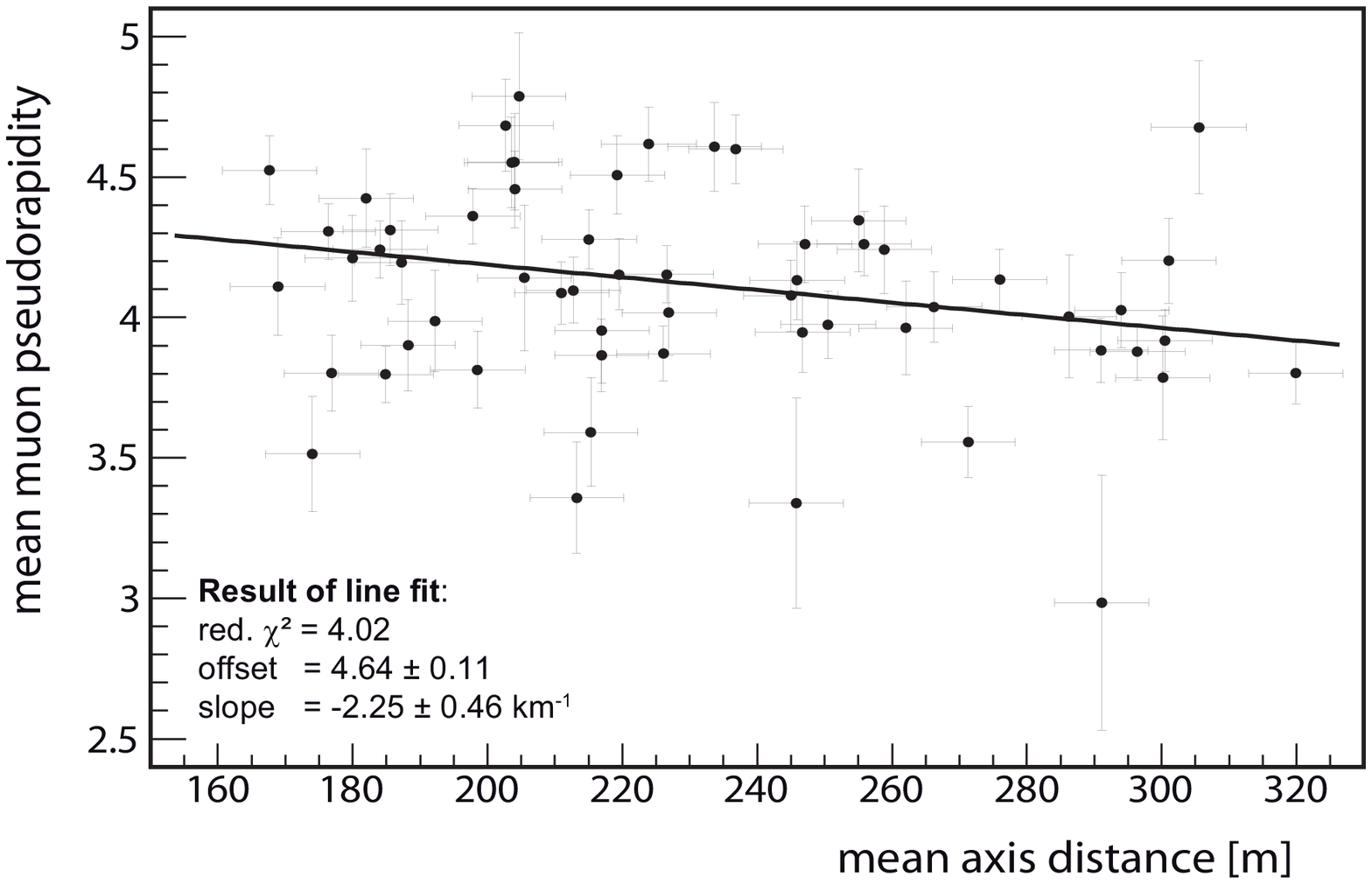}
\includegraphics[width=\columnwidth]{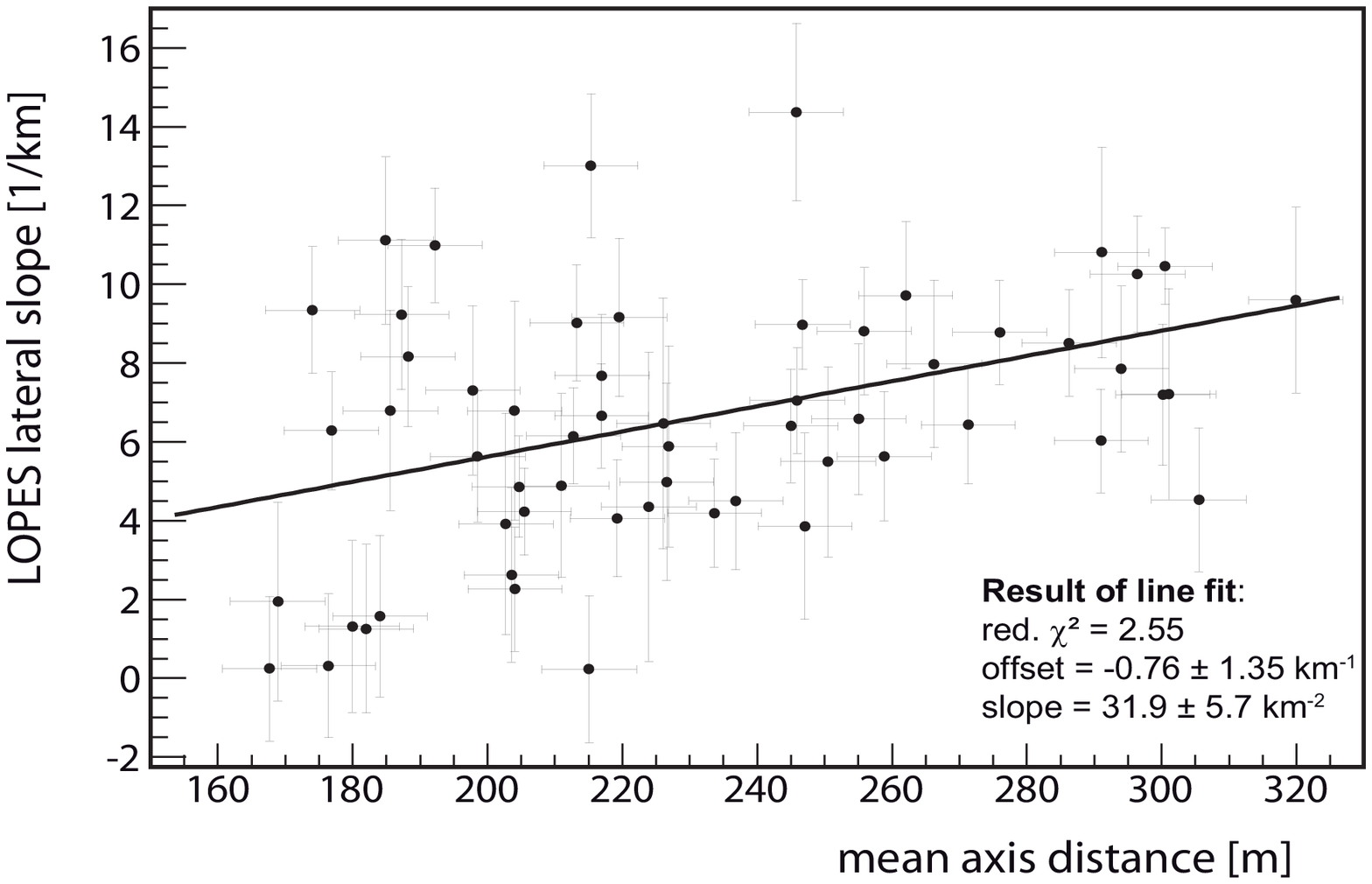}
\includegraphics[width=\columnwidth]{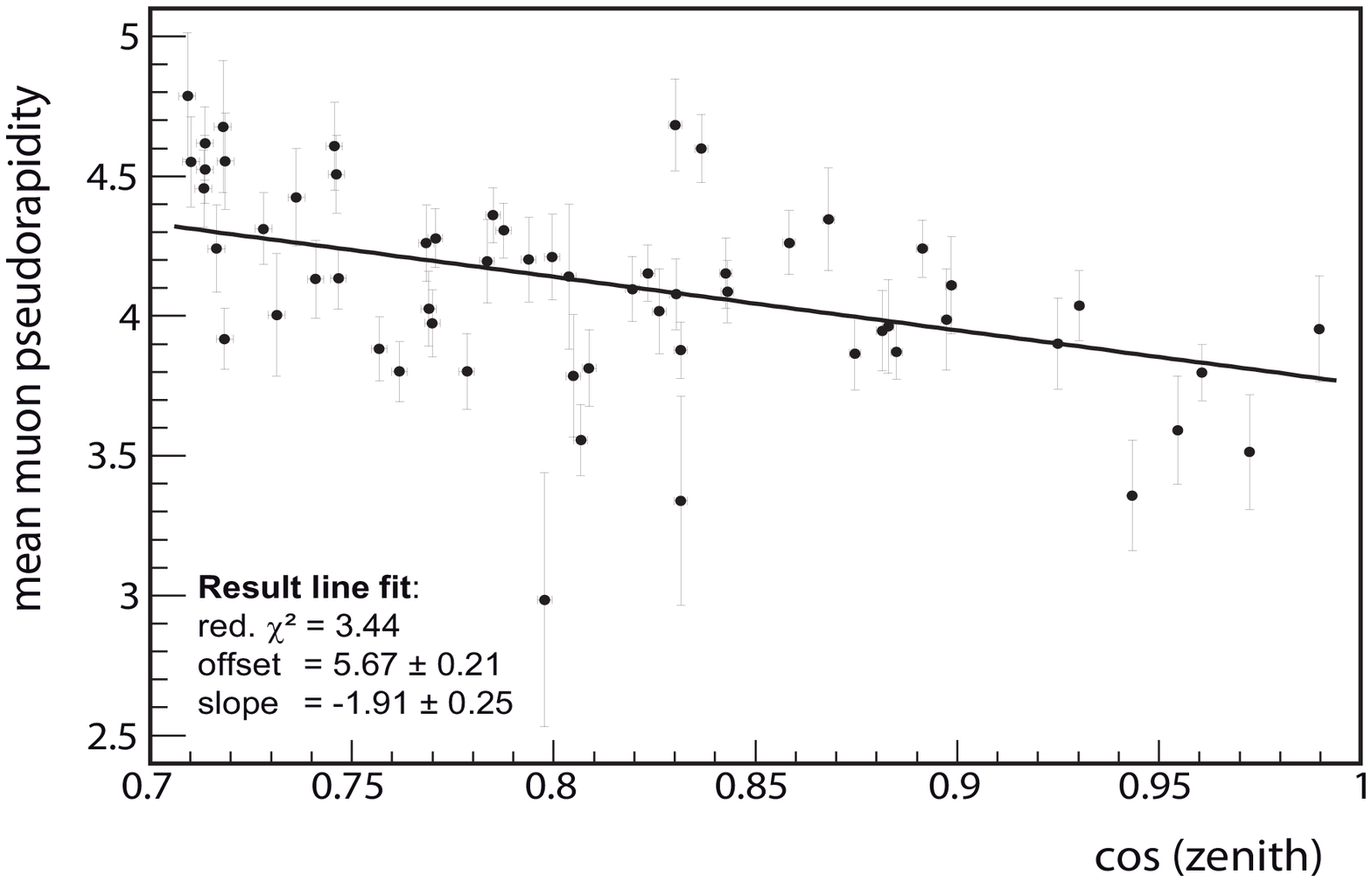}
\includegraphics[width=\columnwidth]{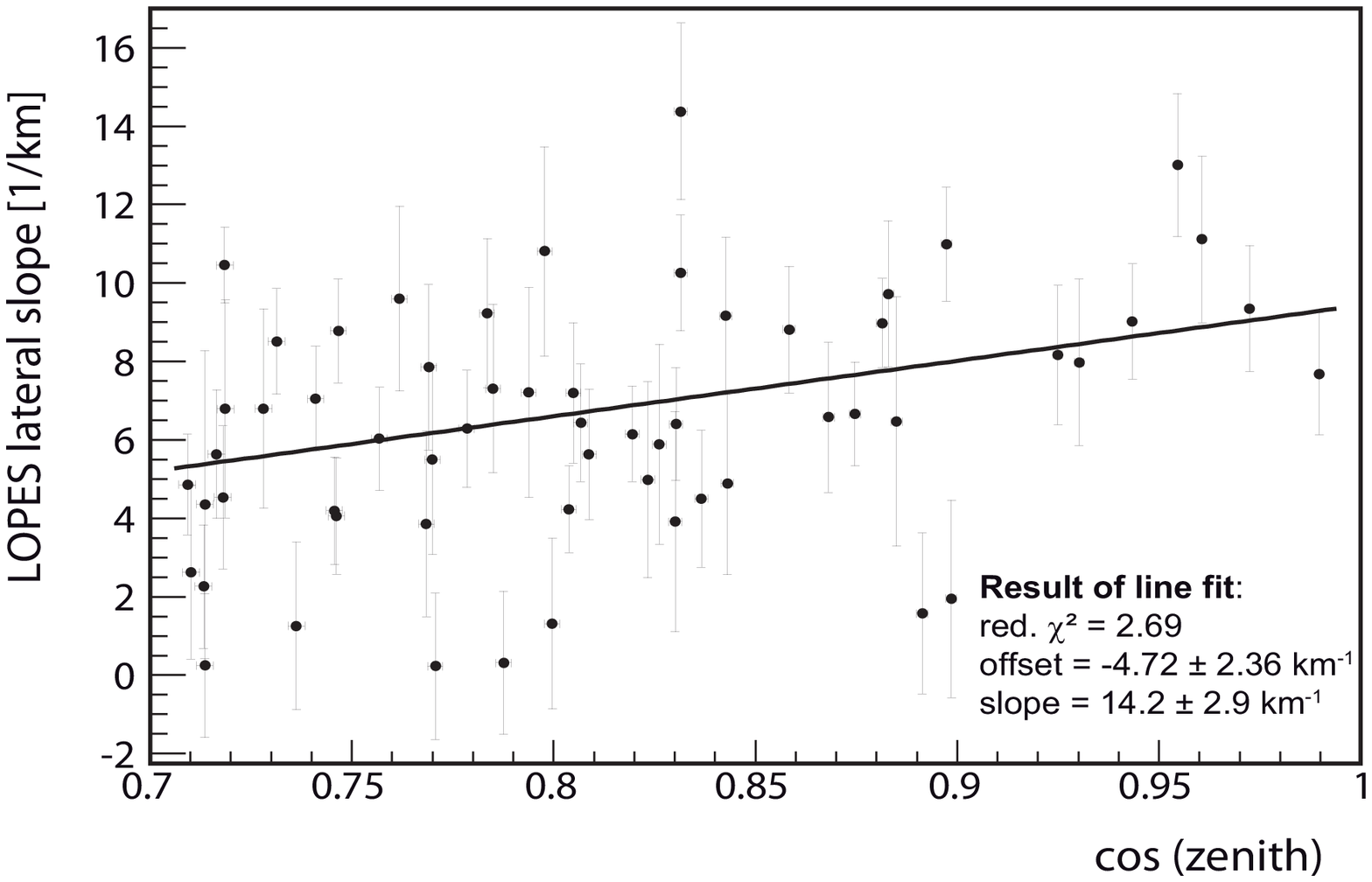}
\caption{Dependencies of the two observables, mean muon pseudorapidity (left) and slope parameter of the radio lateral distribution (right), on the mean distance from the measured muons to the air-shower axis (top) and the shower inclination (bottom). The slopes of the fitted lines are used for correction of the observables.} \label{fig_rmeanAndZenithDependence}
\end{figure*}

The radio lateral distribution is the maximum amplitude of the radio signal in the individual LOPES antennas as a function of the axis-distance of each antenna (figure \ref{fig_radioLDFexample}), where the air-shower radio pulse is identified by cross-correlation beamforming \cite{HuegeARENA_LOPESSummary2010}. We determine the slope of the radio lateral distribution by fitting an exponential function with a slope parameter, where the fit gives us also the statistical uncertainty of the slope parameter. From simulations and theoretical arguments \cite{HuegeUlrichEngel2008} we expect that the slope depends on the geometrical distance of the antenna array to the air-shower cascade, and thus is sensitive to the longitudinal development of the electromagnetic air-shower component and therefore to the primary mass. A steep slope would indicate that the shower developed at a small distance to the radio-detector array. Consequently, we expect an anti-correlation between the mean muon pseudorapidity and the radio-slope parameter, if this theoretical prediction holds true. This is what we indeed observe (figure \ref{fig_correlationPlot}).

The strength of the correlation can be estimated by calculating the correlation coefficient $r$ which is $(-)1$ in case of a perfect (in the absence of measurement uncertainties), linear (anti-)correlation, and 0 in case of no correlation. We calculate $r$ to $-0.64 \pm 0.09$ (fig.~\ref{fig_correlationPlot}). The uncertainty of $r$ has been estimated with the bootstrap method and corresponds to a significance of $7.1 \,\sigma$ for the correlation, i.e.~$r\ge0$ is excluded with $7.1 \,\sigma$. As a consistency check, the significance has also been estimated by fitting a line, minimizing an effective $\chi^2$ taking into account x- and y-uncertainties of each data point \cite{ROOT}. Counting by how many sigmas the slope of the line is different from 0 results in the same significance of $7.1 \,\sigma$ as for the bootstrap method.

The found correlation is either due to a causal link (the common sensitivity of both observables to the shower development), or due to common systematic effects and biases of the measurement devices. In particular, we have checked that the correlation is not only a consequence of known detector effects related to the axis distance and the shower inclination (figure \ref{fig_rmeanAndZenithDependence}). Moreover, possible additional effects due to the geomagnetic angle (= angle between shower direction and geomagnetic field) have been investigated. Although the geomagnetic angle has a large influence on the radio amplitude, this is not true for the slope of the radio lateral distribution. Indeed, we observe no significant effect, neither for the slope parameter nor for the mean muon pseudorapidities. We have removed the systematic axis-distance effect from the correlation by correcting the measured mean pseudorapidity and slope parameter of the radio LDF according to a linear fit to the measured axis-distance dependence. After this correction, the correlation is, as expected, slightly weaker ($r = -0.62 \pm 0.10$), but still has a significance of $6.0 \,\sigma$.

\begin{figure}
\centering
\includegraphics[width=\columnwidth]{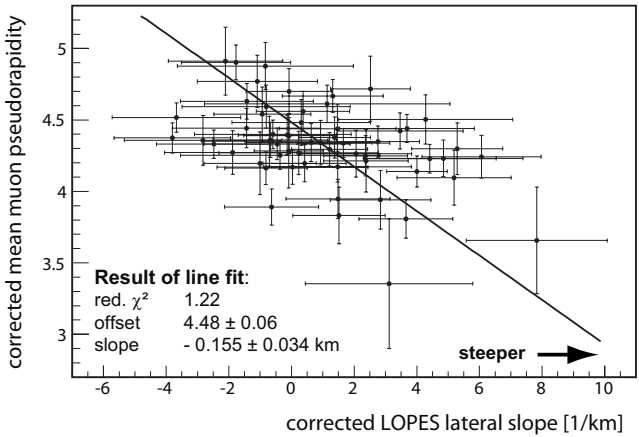}
\caption{Correlation between the mean muon pseudorapidity and slope of the radio lateral distribution after correcting for the distance to the shower axis and the shower inclination.} \label{fig_correlationPlotcorrected}
\end{figure}

The systematic effect of the shower inclination can, in principle, be treated in a similar way, although the physics behind the effect is more complex. Even without any detector effects, we expect a correlation of both observables with the zenith angle, since the geometrical distance between the detector and the development of the air-shower cascade depends not only on the atmospheric depth, but also on the zenith angle $\theta$. Without any model assumptions it is impossible to disentangle detector from physics effects. Thus, we correct both observables for the observed zenith angle dependence as we did for the axis-distance effect. As expected, the correlation (without axis-distance correction) gets weaker after the correction for the shower inclination ($r = -0.56 \pm 0.11$, significance of $5.3 \,\sigma$). When iteratively correcting for both the axis-distance and zenith-angle dependencies, the result stabilizes after a few iterations, and there is still a significant correlation (figure \ref{fig_correlationPlotcorrected}, $r = -0.44 \pm 0.12$, significance of $3.7\,\sigma$).

We investigated whether $|r|<1$ indicates a non-linearity of the correlation between both observables. Therefore, we studied by how much $|r|$ would be reduced due to the measurement uncertainties of both observables if the underlying relation were a perfect linear anti-correlation. In all cases (before and after corrections) the measured value of $r$ is compatible with a linear relation between the mean muon pseudorapidities and the radio-lateral-slope parameter. Hence, more precise measurements or increased statistics will be necessary to test if the correlation is truly or only in first order linear.

We conclude that the observed correlation which remains after the corrections, cannot be explained by known systematic detector effects. Consequently, the correlation is due to the air-shower development in the atmosphere. The fact that the correlation becomes weaker, but is still present after zenith correction, is a strong indication that the slope of the radio lateral distribution is primarily sensitive to the geometrical distance between the antenna array and the air-shower cascade, and not its atmospheric depth in g/cm\textsuperscript{2}. 
The correlation is of particular interest, since the radio emission comes from the electromagnetic component of the air-shower -- mainly due to the geomagnetic deflection of electrons and positrons \cite{KahnLerche1966} and the variation of the net charge during the shower development \cite{Askaryan1962} (see reference \cite{HuegeARENA_ReasVsMGMR2010} for a recent comparison of radio-emission models). Therefore, the correlation is also experimental evidence for the strong link between the development of the electromagnetic and the muonic air-shower component, which itself is a signature of the hadronic component \cite{Andringa2011, DollISVHECRI2010, PintoICRC2011}.
As a consequence, the radio lateral slope must be sensitive -- as the mean pseudorapidity is -- to all parameters which influence the shower development, in particular the energy and mass of the primary particle, and the hadronic interactions. A discussion of those dependencies is, however, beyond the scope of this initial analysis.

To exploit the sensitivity of the air-shower radio emission to the shower development for particle and cosmic-ray physics, quantitative parameters for the shower development like the atmospheric depth of the shower maximum, $X_\mathrm{max}$, have to be reconstructed. This requires additional input by air-shower models or a cross-calibration with another technique, e.g., with fluorescence measurements of $X_\mathrm{max}$ at the Pierre Auger Observatory \cite{FuchsRICAP2011}, or with air-Cherenkov measurements at the Tunka experiment \cite{TunkaRICAP2011}.

At LOPES, a radio reconstruction of $X_\mathrm{max}$ has already been performed using simulations as input \footnote{For the simulations we used REAS3 \cite{LudwigREAS3_2010}, a program which calculates the radio emission of air showers generated by CORSIKA \cite{HeckKnappCapdevielle1998}, which itself is a Monte Carlo tool for a realistic simulation of the air-shower development.}\nocite{LudwigREAS3_2010,HeckKnappCapdevielle1998}, namely the simulated dependencies of $X_\mathrm{max}$ on the radio lateral slope \cite{PalmieriIcrc2011} and the radio wavefront \cite{SchroederIcrc2011}. The $X_\mathrm{max}$ precision of an individual LOPES event is in the order of $100\,$g/cm\textsuperscript{2}, respectively $200\,$g/cm\textsuperscript{2}, when using the first, respectively the second method. The reason for this large uncertainty is the high level of anthropogenic radio noise at the LOPES site. Nevertheless, the simulations indicate that in a situation with negligible noise per-event precisions of better than $30\,$g/cm\textsuperscript{2} can be achieved, i.e.~similar to the precision of the fluorescence and air-Cherenkov technique. To cross-check these simulation-based results, we have estimated the per-event $X_\mathrm{max}$ precision from the correlation of figure \ref{fig_correlationPlotcorrected}. When converting the typical mean muon pseudorapidity of $4$ at a zenith angle of $30^\circ$ and a typical axis distance of $240\,$m to an atmospheric depth, the average per-event uncertainty of the radio lateral slope corresponds to a $X_\mathrm{max}$ uncertainty of about $115\,$g/cm\textsuperscript{2}, which thus is in the same order as the simulation-based results.

Summarizing, a significant correlation has been found between the mean muon pseudorapidity and the slope of radio lateral distributions of individual air-showers. Since the sensitivity of the mean muon pseudorapidity to the longitudinal shower development is established \cite{2011ApelMuonProductionHeights, ZabierowskiIcrc2009, ZabierowskiIcrc2011}, the result serves as first experimental proof that also the radio lateral distribution is sensitive to the longitudinal air-shower development, which implies a sensitivity to the primary mass of high-energy cosmic rays. Since the precision of LOPES is limited by human-made noise, dedicated experimental devices in regions with a low level of ambient noise are required to test whether the sensitivity of radio measurements to the shower development can be used for a reasonable precise reconstruction of $X_\mathrm{max}$ and the mass of the primary cosmic-ray particles.

\begin{acknowledgments}
LOPES and KASCADE-Grande have been supported by the German Federal Ministry of Education and Research. KASCADE-Grande is partly supported by the MIUR and INAF of Italy, the Polish Ministry of Science and Higher Education and by the Romanian Authority for Scientific Research UEFISCDI (PNII-IDEI grant 271/2011). This research has been supported by grant number VH-NG-413 of the Helmholtz Association. We thank Hans Dembinski for fruitful discussions on issues of statistical significances.
\end{acknowledgments}

\bibliography{muonLOPEScorrelationPaper}

\end{document}